\documentclass[3p,times,procedia]{elsarticle}

\usepackage{ecrc}

\volume{00}

\firstpage{1}

\journalname{Physics Procedia}

\runauth{S. Biller}

\jid{phpro}

\jnltitlelogo{Physics Procedia}

\CopyrightLine{2011}{Published by Elsevier Ltd.}

\usepackage{amssymb}

\usepackage[figuresright]{rotating}

\begin{document}

\begin{frontmatter}



\dochead{}

\title{SNO+ with Tellurium}


\author{Steven Biller \\ for the SNO+ Collaboration}

\address{University of Oxford, United Kingdom}

\begin{abstract}
The SNO+ experiment, currently undergoing commissioning, will be a large scale liquid scintillator detector capable of studying a variety of physics topics, with the highest priority being a sensitive search for neutrinoless double beta decay. The collaboration has recently decided to use $^{130}$Te as the candidate isotope for this search, having developed a new approach to the purification and loading of tellurium into liquid scintillator. An initial Phase I demonstrator with a 0.3\% Te loading is expected to reach sensitivities to Majorana neutrino masses approaching the top of the inverted neutrino mass hierarchy. If successful, there is significant scope for further enhancements that could lead to measurements covering the vast majority of the inverted hierarchy range with high sensitivity.
\end{abstract}


\end{frontmatter}



\vskip 0.5in

\section{Introduction}

The SNO+ project will make use of the basic detector infrastructure used for the SNO experiment to produce a large scale liquid scintillation detector capable of exploring a wide range of fundamental physics topics. Situated in the SNOLAB underground laboratory in Sudbury, Canada at a water-equivalent depth below ground of 6800m,  approximately 780 tonnes of liquid scintillator will be viewed by $\sim$9500 photomultiplier tubes (PMTs). The physics targets include a search for neutrinoless double beta decay (0$\nu\beta\beta$), low energy solar neutrinos, geo-neutrinos, reactor neutrinos (for the study of $\Delta m^2_{12}$), supernova neutrinos, and the search for ``invisible modes'' of nucleon decay (which will actually take place during the initial water-filled phase before the transition to scintillator). Numerous modifications and upgrades to the infrastructure have been/are being affected to make this possible, including the installation of a rope net to offset the buoyancy of the scintillator relative to the surrounding water shield, a new cover gas system, a new glovebox for source deployment, scintillator processing systems, upgraded electronics and new calibration systems. 

The highest scientific priority for the experiment is a highly sensitive search for 0$\nu\beta\beta$ decay, the discovery of which would demonstrate lepton number violation and have significant implications for our understanding of neutrino mass as well as potentially providing clues to GUT and leptogenesis models. The original plan for SNO+ was to load $^{150}$Nd into the liquid scintillator for this purpose. However, in the Autumn of 2011, Biller and Chen \cite{BillerChen} began a re-evaluation of the potential for using $^{130}$Te, emphacising several possible advantages:
\begin{itemize}
\item The high 34\% natural abundance of the isotope;
\item The fact that internal U/Th backgrounds can be actively suppressed via Bi-Po $\alpha$ tagging;
\item That external gammas can be sufficiently attenuated in the outer regions of the detector volume;
\item The much lower rate (factor of $\sim$100)  of 2$\nu\beta\beta$ backgrounds relative to $^{150}$Nd;
\item That Te has no inherent atomic absorption in the optical range;
\item That natural tellurium is relatively inexpensive.
\end{itemize}

Initial loading and purification studies were carried out by Yeh {\em et al.} at Brookhaven throughout 2012. The concept subsequently underwent a thorough independent review by the collaboration between August 2012 and February 2013. This resulted in the decision to pursue the Te option as a first priority, which has since been the focus of a full collaboration development effort.

\section{Loading}

The SNO+ scintillator is based on linear alkylbenzene (LAB) as the solvent with 2,5-diphenyloxazole (PPO) as the primary fluor. Initial attempts to load Te in this using carboxylate-based organometallic complexes (as used for Nd) proved unsuccessful in terms of providing a sufficiently clear mixture. However, a new technique \cite{Yeh} was subsequently developed at BNL to produce tellurium-loaded liquid scintillator (TeLS) in which telluric acid is first dissolved in water, in which it is highly soluble. A portion of this mixture is then combined with LAB using an amine-based surfactant to produce a clear and stable ($>$1 year) liquid.

The use of water increases the U/Th content of the mixture relative to pure organic liquids, but the levels are estimated to still be low enough for several percent admixtures assuming similar water activity levels as achieved by SNO ($\sim 10^{-14} - 10^{-15}$ g/g) to allow the use of Bi-Po tagging to reduce associated backgrounds to insignificant levels.

Bench-top spike tests indicate that the U/Th activity levels in the surfactant can be reduced using metal scavenger columns by factors in excess of $\sim$800 in a single pass, suggesting that a multi-pass system should easily be able to achieve purification factors well in excess of the target level, which is estimated to be on the order of $10^4$.

The absence of atomic absorption lines in the wavelength region of PMT sensitivity (Figure 1) also allows for the effective use of secondary wavelength shifters to move the scintillator emission to a region less affected by absorption from the surfactant. Studies suggest an average light level for the full mixture corresponding to between 200 and 300 detected photoelectrons per MeV of deposited energy. This is based on a multi-component scintillator model extrapolating laboratory measurements of absorption, emission and intrinsic light yield of both the full TeLS mixture and individual constituents.  The validity of the model was verified by scaling performance estimates to known scintillator experiments and from an analysis of light production from one litre of scintillator contained in an acrylic container deployed in the SNO detector during 2008.

\begin{figure}[H]
\includegraphics[width=120mm]{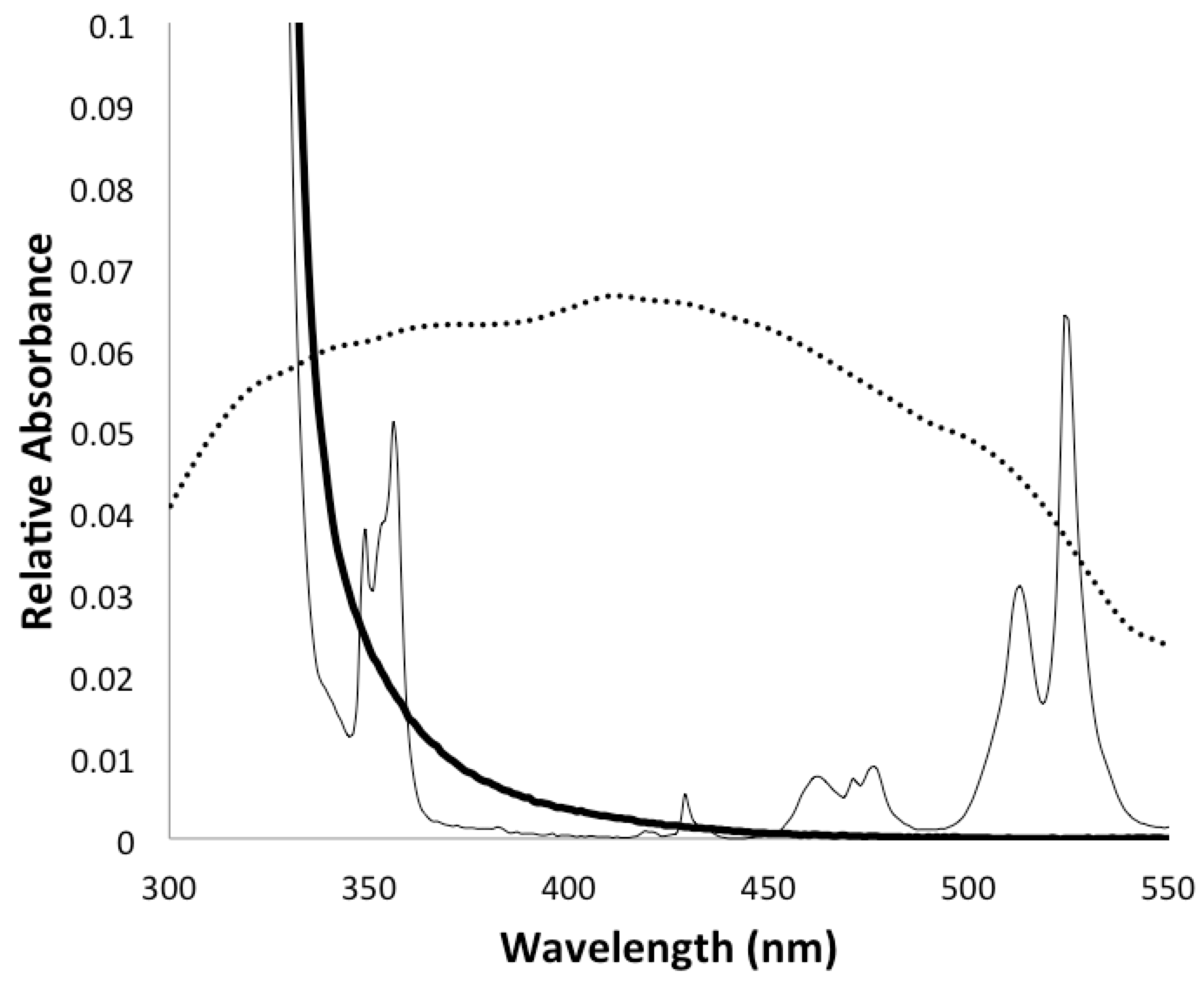}
\caption{Relative absorbance versus wavelength for 0.3\% Te (bold line) and 0.3\% Nd (thin line) loaded scintillator, without secondary wavelength shifters or fluors. The scaled relative PMT response is also shown (dotted line) to indicate the range of wavelength sensitivity for the SNO+ PMTs.}
\end{figure}

\section{Cosmogenics and Purification}

A thorough assessment of potential cosmogenic backgrounds has been undertaken by Lozza {\em et al.} \cite{Lozza} using the ACTIVIA code \cite{ACTIVIA}, cross sections from Silberberg {\em et al.} \cite{Silberberg} and TENDL-2009 database \cite{TENDL} and flux parameterisations from Armstrong and Gehrels \cite{Armstrong}. Variations from using YIELDX code \cite{YIELDX}, TENDL-2012 database, and fluxes from Ziegler \cite{Ziegler} change estimated rates by up to a factor of two. Consistency also checked against CUORE beam activation study \cite{Wang} and the potential KamLAND induced backgrounds.

The conclusion of this study was that reduction factors in excess of $10^4$ are required for the various produced isotopes following surface activation in order to reduce associated backgrounds to insignificant levels. This is approximately the same factor required to for the reduction of the U/Th content in the ``raw'' Te material based on ICP-MS measurements indicating initial levels of $\sim2-3\times10^{-11}$ g/g.

A 2-stage purification procedure has been developed to achieve these goals \cite{purification}. The first of these stages involves two purification passes at surface in which telluric acid is first dissolved in water and then recrystallised using nitric acid, with residual contaminants rinsed off with ethanol. The recrystallisation process is found to be highly tellurium-specific, with the separation of other contaminant resulting in a reduction factor of $\sim$100 or more per pass for all elements so far tested using spikes. 

Following this first stage, several hours are allowed for completion and transport underground, over which time some re-activation will take place. The second stage therefore takes place underground and is intended to address the issue of re-exposure. Here, a less efficient thermal process that is better suited for underground operations will be used in which the telluric acid is dissolved in water at 80$^\circ$C and then cooled to recrystallise the material without further rinsing. Two passes are expected to reduce cosmogenic levels by more than a factor of 100 based on bench-top spike tests, with 50\% of the Te remaining in solution, which will be recycled to the surface system for recovery.

For tellurium mined within a year prior to the telluric acid production, the expectation is that the above procedure combined with a further $\sim$6 months of ``cool-down'' time underground, will result in less than 1 background event from cosmogenic activation in the region of interest per year of operation for a 0.3\% Te loading.

\section{Other Troublesome Backgrounds}

External gamma-ray backgrounds are expected to be dominated by 2.6 MeV $\gamma$'s  from $^{208}$Tl decays in the acrylic vessel and rope net. However, these are attenuated by Compton scattering in the outer region of the detector volume. For a fiducial radius less than $\sim$4m from the centre of the detector, the overall external backgrounds are, in fact, expected to be dominated by $^8$B solar neutrinos.

As for internal backgrounds, the expected levels of U/Th activity can result in a substantial backgrounds near the 0$\nu\beta\beta$ endpoint dominated by supported $^{214}$Bi and $^{212}$Bi decays:

The $^{214}$Bi decay proceeds via $^{214}$Bi$\rightarrow$$^{214}$Po (Q=3.27 MeV) with a branching fraction of 99.979\%. However, these decays can be very efficiently identified by the delayed 7.7 MeV alpha particle emitted by the subsequent decay of $^{214}$Po (t$_{1/2}$=164~$\mu$s) to $^{210}$Pb. This second event will occur outside of the SNO+ detector trigger window of 400~ns in 99.8\% of cases, giving rise to a separate event trigger that can be readily identified. In the other 0.2\% of cases, the alpha will ``pile up'' in the same trigger window as the initial decay. However, this will in most cases lead to a clear distortion in the time distribution of light detected by the PMTs. Another route for the $^{214}$Bi decay to $^{210}$Pb is via $^{210}$Tl, which occurs with a branching fraction of 0.021\%. In this case, a 5.5 MeV alpha from the first step precedes the subsequent decay of $^{210}$Tl (Q=5.49 MeV, half-life = 1.3 minutes). While the longer delay between decays for the latter makes it more difficult to cleanly identify these as coincident events with the same rejection power as for the dominant decay branch, the criteria for this identification is greatly diminished by the much smaller branching fraction.

The $^{212}$Bi decay proceed via  $^{212}$Bi$\rightarrow$$^{212}$Po (Q=2.25 MeV) $\rightarrow$$^{208}$Pb+$\alpha$(8.9 MeV). The last portion of this scheme has an associated half-life of 300ns, meaning that these two decays will pile up within the SNO+ trigger window $\sim$66\% of the time. It is the combined deposition with the $\alpha$ (quenched by a factor of $\sim$10) that produces a potential background in the region of interest. However, as before, such a pile up will generally result in a clear distortion in the time distribution of light detected by the PMTs, allowing for the rejection of such events by a factor greater than 50 (conservatively estimated), which is sufficient to render this contribution insignificant.

\section{Projected Sensitivity and the Pedigree of Bumps}

The basic detector parameters for the Phase I demonstrator are as follows:
\begin{itemize}
\item An effective light level of 200-300 pe/MeV, depending on final optics and choice of secondary shifter.  200 pe/MeV will be assumed here.
\item A loading level of 0.3-0.5\% (0.8-1.3 tonnes $^{130}$Te), depending on final Te system resources.   0.3\% loading will be assumed here.
\item Fiducial Volume: 20-30\%, depending on light level, loading fraction and final backgrounds.  A 20\% fiducial volume will be assumed here (which is still $\sim$10 times the current K-Z fiducial volume).
\end{itemize}

\begin{figure}[H]
\includegraphics[width=150mm]{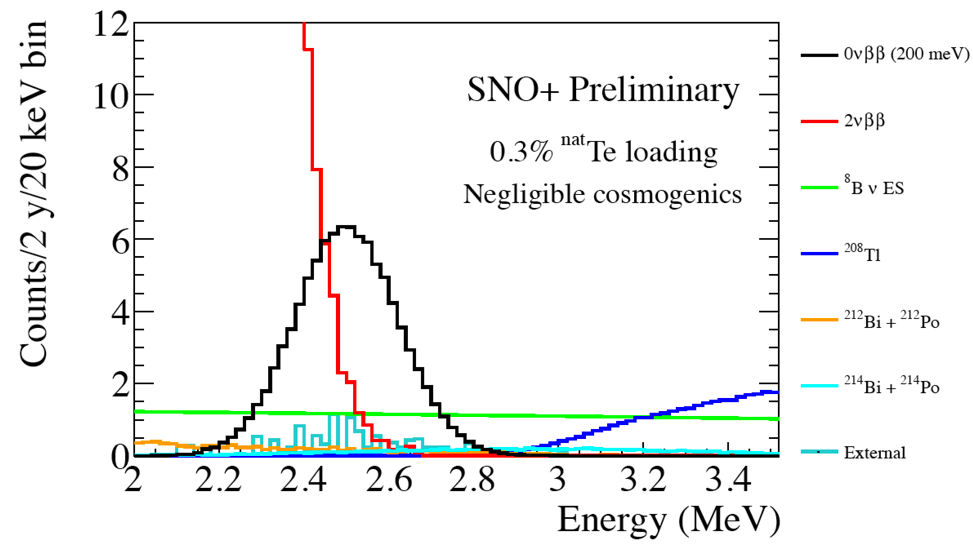}
\caption{Expected average spectra of contribution backgrounds for two live years of data compared to a signal corresponding to a 200meV Majorana neutrino mass, assuming a light level of 200 pe/MeV and a value for the matrix element of 4.}
\label{spectra}
\end{figure}

Figure \ref{spectra} shows the estimated average spectra of contributing backgrounds for two live years of data compared to a signal corresponding to a 200meV Majorana neutrino mass, assuming a value for the matrix element of 4. For a 3-5 year run, the Phase I demonstrator is projected to achieve a Majorana mass sensitivity approaching the top of the inverted neutrino mass hierarchy (depending on the matrix element), comparable to CUORE projections. Higher loadings have been demonstrated, and a future phase with a loading $\sim$10 times higher has the potential to cover the majority of the inverted hierarchy with high sensitivity.

Should a potential signal be observed, an immediate question that arises concerns the extent to which it can be established that the observation is not the result of some unexpected background. There are, in fact, a number of possible handles owing to the advantages of having a large, self-shielded, liquid detector with good spatial resolution and a flexible configuration:
\begin{itemize}
\item Separately measure backgrounds prior to loading and increase or decrease the loading at later stages to see if it scales like a signal;
\item Look at the radial, as well as energy dependence of any potential signal to look for signs of external backgrounds;
\item Look for time dependencies of potential radioactive backgrounds;
\item Remove Te and run it though additional targeted purification systems to reduce/test for any suspected contaminants;
\item If the signal appears to be high enough in mass, could still deploy Nd (or some other isotope) as an independent check;
\item Could upgrade detector with high QE PMTs and better concentrators to improve energy resolution and overall sensitivity;
\item Possible multi-site discrimination using time residuals.
\end{itemize}

\section{Present and Future}

(Updated) The SNO+ detector has begun its commissioning phase, with the initial water-fill to be completed in the Summer of 2014, which will also see the first delivery of telluric acid to SNOLAB. The transition to scintillator will take place in 2015, along with the purification and processing of the telluric acid. Introduction of the isotope will then take place as soon as possible, likely by the end of 2015 or beginning of 2016.

With regard to the potential for further upgrades of the technique, it is worth noting the following:
\begin{itemize}
\item The currently planned Phase I loading is only 0.3\% ; 
\item The current peak SNO PMT efficiency is only $\sim$15\%;
\item The current effective photocathode coverage is only $\sim$50\%;
\item The current limiting external background is from the acrylic vessel;
\item The current fiducial volume is only $\sim$200 tonnes  (relative to the full cavity volume of $\sim$7kt).
\end{itemize}
None of these represent a fundamental limitation. If the Phase I demonstrator performs in line with expectation, there is therefore anticipated to be substantial scope for future development of this approach.

\label{}





\bibliographystyle{elsarticle-num}
\bibliography{<your-bib-database>}



\end{document}